\documentclass[aps,prl,twocolumn,showpacs,superscriptaddress,groupedaddress]{revtex4}
\usepackage{dsfont}
\usepackage{amsthm}
\usepackage{amsmath}
\usepackage{amssymb}
\usepackage{esint}
\usepackage{graphicx}
\usepackage{mathrsfs}             % See geometry.pdf to learn the layout options. There are lots.
\usepackage{graphicx}
\usepackage{amssymb}
\usepackage{epstopdf}
\usepackage{textpos}
\usepackage{lipsum}
\usepackage{color}
\begin{document}
\begin{textblock*}{24cm}(15cm,-1cm)
\fbox{\footnotesize $\substack{\text{MIT-CTP-3881} \\ \\ \text{SU-ITP-17/02}}$}
\end{textblock*} 
\title{Temporal Observables and Entangled Histories}
\date{\today}
\author{Jordan S. Cotler}
\affiliation{Stanford Institute for Theoretical Physics, Stanford University, Stanford, CA 94305}
\author{Frank A. Wilczek}
\affiliation{Center for Theoretical Physics, MIT, Cambridge MA 02139, USA}
\affiliation{Wilczek Quantum Center, Department of Physics and Astronomy, Shanghai Jiao Tong University, Shanghai 200240, China}
\affiliation{Department of Physics, Stockholm University, Stockholm Sweden}
\affiliation{Department of Physics, Arizona State University, Tempe AZ 25287 USA}

\begin{abstract}
We demonstrate that temporal observables, which are sensitive to a system's history (as opposed to its state), implicate entangled histories.  We exemplify protocols for measuring such observables, and algorithms for predicting the (stochastic) outcomes of such measurements.   Temporal observables allow us to define, and potentially to measure, precise mathematical consequences of intrinsically disjoint, yet mutually accessible, branches within the evolution of a pure quantum state.  
\end{abstract}


\maketitle

%%%%%%%%%%%%%%%%%%%%%%%%%%%%%%%%%%%%%%%%%%%%%%%%%%%%%%%%%%%%%%%%%%%%%%%%%%%%%%

Experimenters are achieving new levels of control over the production and manipulation of entanglement among quantum variables.   Much of this effort is inspired by the vision of quantum computing.   But useful quantum computation will require exquisite control of many coupled qubits, and might take several -- perhaps many -- years to achieve.  It could be fruitful to consider other applications of qubit manipulation that are less daunting, and which might also serve as interesting intermediate projects.  In that spirit, here we will describe newly practical protocols that can lead us to an enriched, consequential concept of history in quantum theory.  

{\it Heuristic Example}:  We will work up to more general considerations by starting from a heuristic treatment of a simple but non-trivial example, before supplying precise definitions and protocols.   Consider a standard spin-$\frac{1}{2}$ qubit.  If we aspire to measure quantities which depend upon its history (as opposed to its state) among the simplest formal objects we might consider are temporal operators like
\begin{eqnarray}
\label{ABeq}
A ~&\equiv&~ \sigma_2 \odot \sigma_1 \nonumber \\
B ~&\equiv&~ \sigma_1 \odot \sigma_3
\end{eqnarray}
which act on history state space $\mathcal{H}_{t_2} \odot \mathcal{H}_{t_1}$, the tensor product of the ordinary state spaces at times $t_1, t_2 > t_1$.  Here ``$\odot$" denotes a tensor product \textit{in time} between the Hilbert spaces $\mathcal{H}_{t_1}$ and $\mathcal{H}_{t_2}$ \cite{griffiths}.

Note that $A$ and $B$ commute, though their component factors at times $t_1$ and $t_2$ both anticommute.  Thus it is plausible that $A$ and $B$ can be measured simultaneously.  Indeed, it is easy to verify that they have the simultaneous eigenhistories:
\begin{eqnarray}
\label{eigenHists1}
|v_{++}\rangle ~&=&~  1/2\,(\, | \uparrow \uparrow \rangle \, + \,  | \downarrow \uparrow \rangle \, + \, i | \downarrow \downarrow \rangle \, - i \, | \uparrow \downarrow \rangle \,)\nonumber \\
|v_{+-}\rangle ~&=&~ 1/2\,(| \uparrow \uparrow \rangle \, - \,  | \downarrow \uparrow \rangle \, + \, i | \downarrow \downarrow \rangle \, + i \, | \uparrow \downarrow \rangle\,) \nonumber \\
|v_{-+}\rangle ~&=&~ 1/2\,(\,| \uparrow \uparrow \rangle \, + \,  | \downarrow \uparrow \rangle \, - \, i | \downarrow \downarrow \rangle \, + i \, | \uparrow \downarrow \rangle\,) \nonumber \\
|v_{--}\rangle ~&=&~ 1/2\,(\,| \uparrow \uparrow \rangle \, - \,  | \downarrow \uparrow \rangle \, - \, i | \downarrow \downarrow \rangle \, - i \, | \uparrow \downarrow \rangle\,)
\end{eqnarray}
We note that each of these eigenhistories is entangled \cite{entangledHistories}, in the sense that it cannot be written as a product of a state in $\mathcal{H}_{t_2}$ and a state in $\mathcal{H}_{t_1}$. 
\vskip0in
In what sense are temporal operators such as $A, B$ observables?   That is not completely obvious, since (for example) measuring the first component of $A$, i.e. $\sigma_1 (t_1)$, disrupts the evolution of our qubit.  Speaking loosely, it collapses the wave function, after which our eigenhistories lose any obvious relevance.  Sequential measurements of this kind are not uninteresting, of course (see, especially, \cite{leggettGarg, Aharonov1}), but they do not offer direct access to intrinsic properties of the system's undisturbed evolution, as we would like true temporal observables to do.  Standard correlation functions access weighted expectation values of temporal observables, but not their full operator structure.

\begin{figure}
\centering
\includegraphics[scale=0.25]{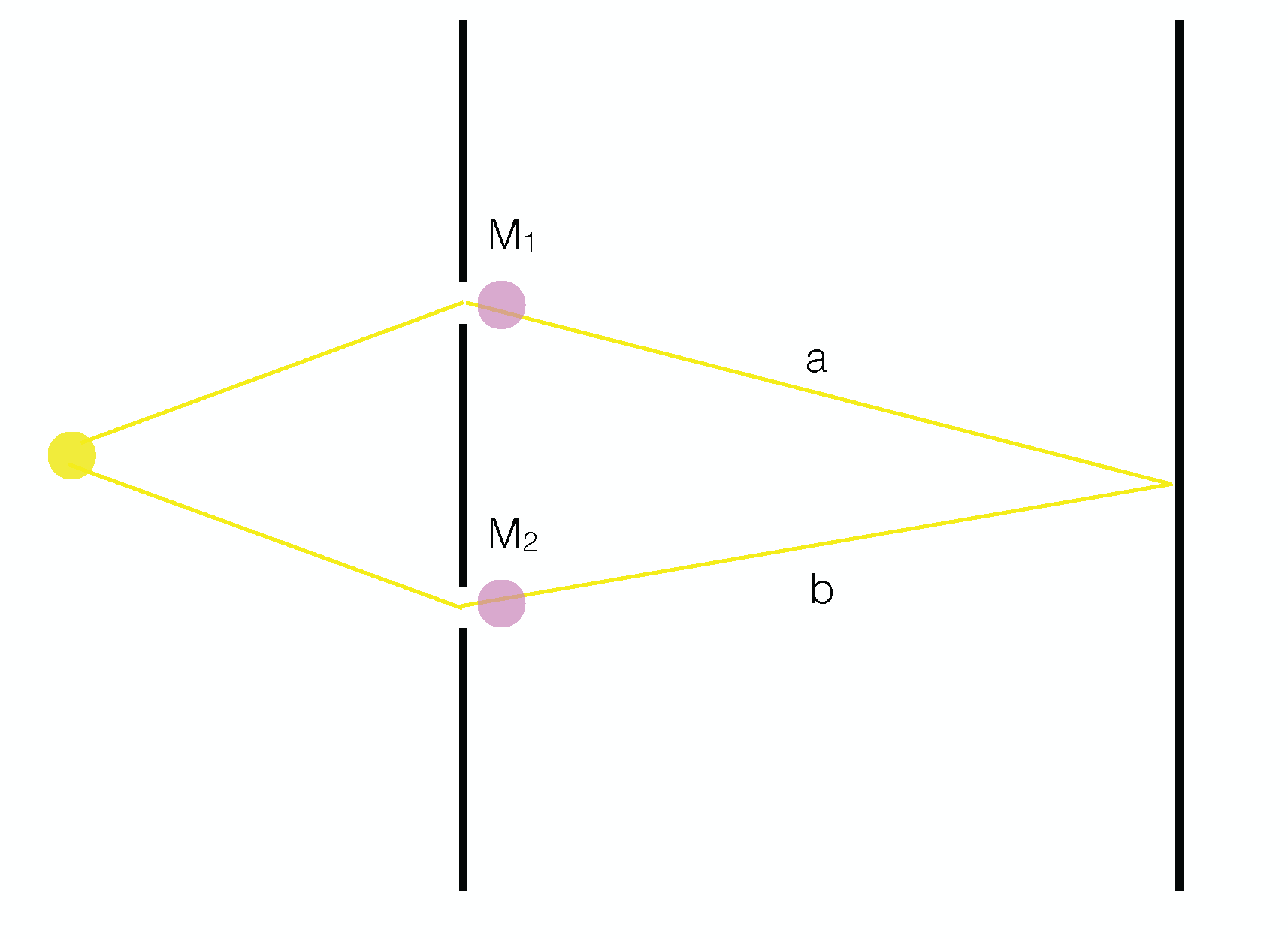}
\caption{Two-slit experiment with monitor bits.} 
\label{fig:twoSlit}
\end{figure}

{\it Monitor Bits; Monitored Two-Slit Experiment}:  Our protocol for observing $A, B$ involves introducing auxiliary ``monitor'' qubits which correlate with, but do not observe, the behavior of our system at $t_1$ and $t_2$.  (Related constructions appear in discussions of model detectors in Griffiths' book \cite{griffiths} and in discussions of ``multiple-time states" \cite{Aharonov1, Aharonov2, Aharonov3}.)  We can illustrate the essence of our construction using a variation on the classic two-slit experiment which is often used to exhibit the central novelties of quantum mechanics.   It will be sufficient to adopt some standard idealizations: two-dimensional geometry, point ``slits'', fixed polarization.   See Figure \ref{fig:twoSlit}.

The novel feature of our set-up is the inclusion of two monitor qubits, one for each slit.   The monitor qubit $M_1$ for slit 1 begins in the down state, and flips if and only if the photon passes through slit 1.   More precisely, its dynamics are governed by the unitary transformation
\begin{eqnarray}
|{\rm Y} \rangle \otimes | \downarrow \rangle ~&\rightarrow~ \, |{\rm Y} \rangle \otimes | \uparrow \rangle \nonumber \\
|{\rm N} \rangle \otimes | \downarrow \rangle ~&\rightarrow~ \, |{\rm N} \rangle \otimes | \downarrow \rangle \nonumber \\
|{\rm Y} \rangle \otimes | \uparrow \rangle ~&\rightarrow~ \, |{\rm Y} \rangle \otimes | \downarrow \rangle \nonumber \\
|{\rm N} \rangle \otimes | \uparrow \rangle ~&\rightarrow~ \, |{\rm N} \rangle \otimes | \uparrow \rangle  
\end{eqnarray}
where Y indicates that yes the photon has passed through, while N indicates that no it has not.   (Griffiths uses a similar construction of model ``detectors'' in several of the thought-experiments described in \cite{griffiths}.)   This is the evolution of a CNOT gate, with the photon's passage acting as the control.  

The monitor bit $M_2$ for slit 2 is, of course, quite similar, but it flips if and only if the photon passes through slit 2.  Note that in neither case do we perform an irreversible measurement, so to say the photon ``passes through'', though standard and convenient, is a bit loose.  

Now if we proceed to measure the $\hat z$ component of either one of the auxiliary spins, prior to the arrival of the photon at the screen, we will not observe a nontrivial interference pattern.  If the amplitudes for propagation from slits 1, 2  to our observing point are $a, b$, the probability for arrival will be governed by $|a|^2$ if we observe $M_1$ to be spinning up or $M_2$ to be spinning down, and by $|b|^2$ in the opposite cases.

On the other hand, if we measure the total spin of the auxiliary system $M_1 \otimes M_2$, we will observe interference.  Indeed, we will find two different patterns, depending on the value of that total spin.  If the total spin is 1, corresponding to the symmetric state of the spins, we will observe $\frac{1}{2} |a + b|^2$.  If the total spin is 0, corresponding to the antisymmetric state of the spins, we will observe $\frac{1}{2} |a - b|^2$.  Analogous operations will allow us to measure $A$ without measuring $\sigma_1(t_1)$ or $\sigma_2(t_2)$ separately.

We can also perform other operations on the auxiliary bits, such as phase-shifting one of them.  We can also, in principle, measure peculiar combinations, say $\vec s_1 \cdot \vec s_2 + \beta s^x_1$, and generate unusual interference patterns.   We can do these measurements after the photon has passed by the slits, (but before it arrives at the screen).  We can also consider the effect of measuring arrival at the screen on the state of the monitor bits.  This is a way of producing entangled pairs (i.e., EPR or Bell pairs).  Indeed, if the photon arrives at a position corresponding to a zero in the interference pattern for total monitor spin 1, then the total spin of the monitor bits must be 0, and {\it vice versa}.  More generally, if the contributions of the photon amplitude at a given point on our screen from the two slits due to paths emanating from the two slits are  $a$, $b$ respectively, as before, then we find
\begin{equation}
\pi ~=~ \left(\begin{array}{ccc} |a|^2 & a b^* \\ b a^* & |b|^2\end{array}\right)
\end{equation}
for the (unnormalized) density matrix of the monitor bits in the basis $\{|\uparrow\,\uparrow\rangle, |\uparrow\,\downarrow\rangle, |\downarrow\,\uparrow\rangle, |\downarrow\,\downarrow\rangle \}$.

%%%%%%%%%%%%%%%%%%%%%%%

{\it From Monitor Bits to Temporal Observables}: With these preparations, we are ready to discuss our protocol for monitoring the time evolution of a single system.  For simplicity, consider a system of a single spin-$\frac{1}{2}$ particle, evolving from time $t_1$ to time $t_2$ with unitary $U$.

Our goal will be to understand the time evolution of the single spin system with respect to a preferred choice of orthonormal basis at each instant in time.  In particular, suppose at time $t_1$ we choose a ``preferred" orthonormal basis $\mathcal{B}_1=\{|a\rangle, |\overline{a}\rangle\}$ and at time $t_2$ we choose $\mathcal{B}_2=\{|b\rangle, |\overline{b}\rangle\}$.  If we are given an initial state of our system $|s_1\rangle = \alpha \, |a\rangle + \beta \, |\overline{a}\rangle$, then we consider the history
\begin{align}
|\Psi_{\mathcal{B}_1, \mathcal{B}_2}\rangle =& \, \mathcal{A}(a\to b)\,\alpha \,|b\rangle\odot|a\rangle + \mathcal{A}(a\to \overline{b})\,\alpha\, |\overline{b}\rangle\odot|a\rangle \nonumber\\
& \, + \, \mathcal{A}(\overline{a}\to b)\,\beta \,|b\rangle\odot|\overline{a}\rangle + \mathcal{A}(\overline{a}\to \overline{b})\,\beta \,|\overline{b}\rangle\odot|\overline{a}\rangle
\end{align}
where $\mathcal{A}(a\to b) \equiv \langle b | U |a\rangle$ is the amplitude to transition from $a$ to $b$ (and similarly for the other terms), and again the ``$\odot$" denotes a tensor product \textit{in time} between the Hilbert spaces $\mathcal{H}_{t_1}$ and $\mathcal{H}_{t_2}$.  Note that if we choose $\mathcal{B}_1, \mathcal{B}_2$ so that $|a\rangle = |s_1\rangle$ and $|b\rangle = |U s_1\rangle$ in evident notation, the above history state simply becomes $|U s_1\rangle|s_1\rangle$.  This makes sense, since in this case our bases $\mathcal{B}_1, \mathcal{B}_2$ are tracking the unitary evolution. 

To access $|\Psi_{\mathcal{B}_1, \mathcal{B}_2}\rangle$ in an experiment, suppose that we have two monitor bits that we couple to the main system.  At time $t_1$, we apply a controlled unitary gate which makes the first monitor qubit $|a\rangle$ if the state of the main spin is $|a\rangle$, and $|\overline{a}\rangle$ if it is $|\overline{a}\rangle$.  At time $t_2$, after the unitary time evolution $U$ has been applied to the main system, we apply a controlled unitary gate which makes the second monitor qubit $|b\rangle$ if the state of the main spin is $|b\rangle$, and $|\overline{b}\rangle$ if it is $|\overline{b}\rangle$.  Then the final state of the system is
\begin{align}
& \mathcal{A}(a\to b)\,\alpha \,|b\rangle_{\text{main}}|b\rangle\otimes|a\rangle + \mathcal{A}(a\to \overline{b})\,\alpha\, |\overline{b}\rangle_{\text{main}}|\overline{b}\rangle\otimes|a\rangle \nonumber\\
+ \, & \mathcal{A}(\overline{a}\to b)\,\beta \,|b\rangle_{\text{main}}|b\rangle\otimes|\overline{a}\rangle + \mathcal{A}(\overline{a}\to \overline{b})\,\beta \,|\overline{b}\rangle_{\text{main}}|\overline{b}\rangle\otimes|\overline{a}\rangle
\end{align}
where we use ``$\otimes$" to emphasize a tensor product between the auxiliary qubits which is \textit{not} temporal.
Projecting the main system onto $\frac{1}{\sqrt{2}}(|b\rangle + |\overline{b}\rangle)$ and then tracing it out, the monitor qubits exactly equal the history state $|\Psi_{\mathcal{B}_1, \mathcal{B}_2}\rangle$, but with $\otimes$'s instead of $\odot$'s.  This is the main point: the monitor qubit procedure allows us to access a superposition of records of the \textit{history} of our system, which is isomorphic to a ``history state" by switching out $\otimes$'s with $\odot$'s.

\textit{Entangling Our Knowledge of the Past:} To be very concrete, let us consider an example.  Suppose we start with a particle in state $|z^+\rangle$, and let $\mathcal{B}_1 = \{|x^+\rangle = \frac{1}{\sqrt{2}}(|z^+\rangle + |z^-\rangle), |x^-\rangle = \frac{1}{\sqrt{2}}(|z^+\rangle - |z^-\rangle)\}$ and also $\mathcal{B}_2 = \{|z^+\rangle, |z^-\rangle\}$.  We further suppose that the unitary time evolution of the main system is trivial, namely $U = \textbf{1}$.  Then after performing the auxiliary qubit procedure, projecting the main system onto $\frac{1}{\sqrt{2}}(|z^+\rangle + |z^-\rangle)$ and tracing it out, we obtain the state of the monitor qubits
\begin{equation}
\label{monitorEq}
\frac{1}{2} |z^+\rangle \otimes |x^+\rangle + \frac{1}{2} |z^+\rangle \otimes |x^-\rangle + \frac{1}{2} |z^-\rangle \otimes |x^+\rangle - \frac{1}{2} |z^-\rangle \otimes |x^-\rangle\,.
\end{equation}
Now suppose we measure the state of the monitor qubits in the $\mathcal{B}_2 \otimes \mathcal{B}_1$ basis, and that we read out $|z^+\rangle \otimes |x^-\rangle$ (which we will get with probability $|1/2|^2 = 1/4$).  Then the world as it currently is must be consistent with the particle having been $|x^+\rangle$ at time $t_1$ and $|z^-\rangle$ at time $t_2$.  This is similar to the double-slit experiment from before, in which collapsing the state of the monitor qubits picks out a particular version of the past.  Our setup is a generalization in the sense that we are picking out a \textit{sequence} of events, whereas in the double-slit case we picked out a \textit{single} event (namely, the passage of the particle through one or both slits).

Now suppose that we  measure instead the state of the monitor qubits in Eqn. (\ref{monitorEq}) in an \textit{entangled} basis, such as one containing $\frac{1}{\sqrt{2}}(|z^+\rangle \otimes |x^+\rangle - |z^-\rangle \otimes |x^-\rangle)$. If upon measurement we read out $\frac{1}{\sqrt{2}}(|z^+\rangle \otimes |x^+\rangle - |z^-\rangle \otimes |x^-\rangle)$ (which will happen with probability $1/2$), then our best description of the past is an entangled history.  In other words, a superposition of \textit{sequences} of events is consistent with the current state of our world, and best describes our knowledge about the past of our world.

As in our double-slit experiment, we have the flexibility, by selecting the basis in which we measure the monitor qubits, to choose how to ``collapse the past."  For example, we can measure the state of the monitor qubits using the operators $A$, $B$ given in Eqn. \!(\ref{ABeq}), thereby forcing the best description of the past to be one of the entangled eigenstates given in Eqn. \!(\ref{eigenHists1}).

{\it Discussion}: 1. In the preceding examples we have taken a non-interacting spin as our dynamical system of interest, but as explained above our protocols for defining temporal observables work much more generally.  Indeed, any unitary evolution of the spin is allowed.  Furthermore, our approach generalizes to multiple spins or other degrees of freedom.  In this context, it is natural to consider spatiotemporal observables, associated to space-time events, which bring in two distinct spins located at $(x_2, t_2)$ and $(x_1, t_1)$.  These are the operator forms of correlation functions (expectation values) whose arguments depend on space and time.  Here we are no longer dealing with the history of a particular particle, but rather sampling the history of a field variable.

2.  History space figures prominently in the foundational work of Robert Griffiths and many investigations based upon it \cite{Griffiths1, Griffiths2, griffiths}.  (It is also implicit in the important, independent approach pursued by Aharonov {\it et al}. \cite{Aharonov1}.)   Our use of it here is rather different and more fine-grained, however. Griffiths' formalism introduces a degenerate inner product, based on his $K$ operator.   Probabilities are calculated using this $K$ operator.  Griffiths' approach is adapted to observations which can be made using ordinary states at the initial and final times, and allows only limited or indirect inferences about the detailed evolution of dynamical variables at intermediate times.   One sign of this is that when applied to a system with $n$ degrees of freedom it yields, after removing states of zero norm, a Hilbert space of dimension $n^2$ regardless of how many intermediate times are involved \cite{Diosi1}; another is the difficulty  of rendering temporal operators observable.   Here, by expanding our consideration to include monitor bits and their measurement,  we gain access to many more parameters.  We can resolve the entire history space, using a complete set of commuting (temporal) observables and employing the canonical inner product, as exemplified above. 

3.   An alternative approach to accessing history space employs multiple copies of the initial state.  It is very smooth mathematically, but more challenging to implement physically, and its interpretation is less direct.  

With the benefit of hindsight we will start by defining the action of $A$ on a history that evolves from a two-spin product state $|s_b, s_a \rangle$ at an early time.   We introduce two monitor bits, one which monitors the 2-component of spin  $s_a$ at time $t_1$, and another which monitors the 1-component of spin $s_b$ at time $t_2$.  In an evident notation, the entire system evolves as 
\begin{eqnarray}
~&{}&~ \!\!\!\!\!\!\!\!| s_b , s_a , \downarrow \downarrow \rangle \nonumber \\
&{}& \rightarrow  | s_b , \frac{1 + \sigma_1}{2} s_a , \downarrow \uparrow \rangle + | s_b , \frac{1 - \sigma_1}{2} s_a , \downarrow \downarrow \rangle  \nonumber \\
&{}& \rightarrow  | \frac{1 + \sigma_2}{2} s_b , \frac{1 + \sigma_1}{2} s_a , \uparrow \uparrow \rangle
+ | \frac{1 - \sigma_2}{2} s_b , \frac{1 + \sigma_1}{2} s_a , \downarrow \uparrow \rangle \nonumber \\
&{}& \,\,\,+ | \frac{1 + \sigma_2}{2} s_b , \frac{1 - \sigma_1}{2} s_a , \uparrow \downarrow \rangle
+ | \frac{1 - \sigma_2}{2} s_b , \frac{1 - \sigma_1}{2} s_a , \downarrow \downarrow \rangle \nonumber \,.\\
~&{}&~
\end{eqnarray}
Now if we project on the entangled monitor bit state which corresponds to the product of spins being 1, i.e. $| \uparrow \uparrow \rangle + | \downarrow \downarrow \rangle$ (or, alternatively stated, measure the value of $\sigma_3 \otimes \sigma_3$ to be +1) we arrive at 
\begin{eqnarray}\label{centralEquation}
&{}& | \frac{1 + \sigma_2}{2} s_b , \frac{1 + \sigma_1}{2} s_a \rangle + | \frac{1 - \sigma_2}{2} s_b , \frac{1 - \sigma_1}{2} s_a \rangle \nonumber \\
&=&  \frac{1 + \sigma_2 \otimes \sigma_1}{2}  | s_b, s_a \rangle\,.
\end{eqnarray}
Similarly, projecting on the entangled monitor bit state  $| \downarrow \uparrow \rangle + | \uparrow \downarrow \rangle$ produces $\frac{1 - \sigma_2 \otimes \sigma_1}{2}  | s_b, s_a \rangle$.   Thus appropriate measurements on the monitor bits project on the eigenhistories of $A$, thereby implementing $A$ as an observable on $|s_b, s_a\rangle$.  

In the special case $s_a = s_b = s$, our protocol produces the desired action of $A$ on the history associated with the Schr\"odinger evolution of $s$, allowing us to access $A$ applied to that Schr\"odinger evolution.  There is an isomorphism between diagonal elements of the tensor product Hilbert space and the history Hilbert space of an individual system.   

It is straightforward to demonstrate that if we allow the factors in $B$ and their corresponding monitor bits to operate infinitesimally later than those in $A$, {\it e.g.} letting its $\sigma_3$ and the corresponding monitor bit
act at $t_1 + \epsilon$, then projecting on the $\pm$ eigenstates of the product of the monitor bits, as we did for $A$, results in a decomposition of the Schr\"odinger history of $s$ into precisely the entangled histories which appeared in our heuristic discussion, according to
\begin{eqnarray}
&{}& \!\!\!\!\!\!\!\!\!\!\!\!| s(t_2), s(t_1) \rangle \nonumber \\ 
&=& ( \frac{1 + \sigma_1 \otimes \sigma_3}{2} \, \frac{1 + \sigma_2 \otimes \sigma_1}{2} +  \frac{1 - \sigma_1 \otimes \sigma_3}{2} \, \frac{1 + \sigma_2 \otimes \sigma_1}{2} \nonumber \\
&&\quad + \, \frac{1 + \sigma_1 \otimes \sigma_3}{2} \, \frac{1 -  \sigma_2 \otimes \sigma_1}{2} \nonumber \\
&&\quad + \, \frac{1 - \sigma_1 \otimes \sigma_3}{2} \, \frac{1 - \sigma_2 \otimes \sigma_1}{2} )  |s(t_2), s(t_1) \rangle  \nonumber \\
&=& v(++) \langle v(++) |s(t_2), s(t_1) \rangle + v(-+) \langle v(-+) |s(t_2), s(t_1) \rangle \nonumber \\
&&\quad +\, v(+-) \langle v(+-) |s(t_2), s(t_1) \rangle \nonumber \\
&&\quad +\, v(--) \langle v(--) |s(t_2), s(t_1) \rangle\,.
\end{eqnarray}

The absolute squares of the coefficients give the probabilities for the Schr\"odinger evolution of $s$ to have run through the corresponding history.  Writing $s = (\cos \frac{\theta}{2} e^{i \phi/2} , \sin \frac{\theta}{2} e^{-i \phi/2})^{\rm T}$, we find, for example
\begin{eqnarray}
&{}&\!\!\!\!\!\!\!\!\!\!\!\!\!\!\!\!\!\!\!\!| \langle v(++) |s(t_2), s(t_1) \rangle |^2 \nonumber \\ 
&=& \frac{1}{4} \, + \, \frac{\sin \theta}{8} (\cos \theta (\cos \phi + \sin \phi) + \frac{1}{2} \sin \theta \sin 2 \phi) \nonumber \\
&{}&
\end{eqnarray}
This specific result has no special importance, beyond predicting the result of a particular class of experiments, but it serves to demonstrate how our treatment of temporal observables leads to specific, quantitative, testable consequences.  Note that the probability here computed is sensitive to interference between different parts of our entangled history.

As was the case for the monitored two-slit experiment, here too we can consider a dual procedure, whereby different measurements on the final state of our two spins yield different, but predictable, density matrices for the monitor bits.

%There is no contradiction, however: The $K$ operation and the associated inner product, acting on an expanded space which includes tensor inputs and monitor bits and their histories (as opposed to just the original system's histories) also generates the probabilities for observable transitions.  

4. When our information about a system is summarized in an entangled history, we have discovered facts about its past which cannot be captured by saying that it passed through a specific temporal sequence of states.  Rather we infer, as our best description, its parallel evolution through several distinct sequences \cite{entangledHistories, Nowakowski1}.  In common language, we would say that several distinct patterns of development took place in parallel.  Thus entangled histories are a tangible mathematical reflection of the ``many worlds'' interpretation of quantum theory.  To render temporal entanglement observable, as a practical matter, we must focus on very small worlds, that haven't diverged very much, and identify effects of interference among them.  We can do that using temporal operators based on small numbers of monitor bits, as we have exemplified above.   

% 2. relation to Griffiths
% 1. from correlation functions to observable operators, generally
% 4. relation to many worlds interpretation
% 3. [Jordan] relativistic QFT, Wheeler-deWitt (?)

\textbf{Acknowledgements.\quad}JC is supported by the Fannie
and John Hertz Foundation and the Stanford Graduate Fellowship program.  FW is supported by the U.S. Department of Energy under contract No. DE-FG02-05ER41360 and by the Swedish Research Council Grant 335-2014-7424.

%%%%%%%%%%%%%%%%%%%%%%%%%%%%%%%%%%%%%%%%%%%%%%%%%%%%%%%%%%%%%%%%%%%%%%%%%%%%%%

\end{document}